\documentclass[apjl]{emulateapj}
\usepackage{graphicx,natbib,apjfonts}

\begin{document}

\date{\today}
\title{Evidence for merging or disruption of red galaxies from the evolution
of their clustering}
\author{Martin White${}^1$}
\author{Zheng Zheng${}^{2,3}$}
\author{Michael J. I. Brown${}^{4,5}$}
\author{Arjun Dey${}^6$}
\author{Buell T. Jannuzi${}^6$}
\affiliation{${}^1$Departments of Physics and Astronomy,
University of California, Berkeley, CA 94720}
\affiliation{${}^2$Institute for Advanced Study, Princeton, NJ 08540}
\affiliation{${}^3$Hubble Fellow}
\affiliation{${}^4$Princeton University Observatory, Peyton Hall,
Princeton, NJ 08544-1001}
\affiliation{${}^5$H.N. Russell Fellow}
\affiliation{${}^6$National Optical Astronomy Observatory,
Tucson, AZ 85726-6732}

\begin{abstract}
The formation and evolution of massive red galaxies form a crucial
test of theories of galaxy formation based on hierarchical assembly.
In this letter we use observations of the clustering of luminous red
galaxies from the Bo\"{o}tes field and N-body simulations to argue that
about $1/3$ of the most luminous satellite galaxies appear to undergo
merging or disruption within massive halos between $z\simeq 0.9$ and
$z\simeq 0.5$.
\end{abstract}

\maketitle

\section{Introduction}

The assembly of the most massive galaxies is a key test of cold dark matter
(CDM) models of galaxy formation, as the ongoing growth of massive galaxies
via mergers is a generic feature of  hierarchical CDM models.
Observationally, the most massive galaxies have little ongoing star formation,
and the bulk of their stellar mass was formed at $z>2$
\citep[e.g.][and references therein]{BowLucEll,Tra00,Coo06}.
If there is appreciable growth of these galaxies at $z<1$, this must be 
due to galaxy mergers, as predicted by the hierarchical CDM models.

Evidence for the ongoing assembly of massive galaxies is inconclusive.
While the stellar mass within the red galaxy population has doubled since
$z=1$ \citep{Bel04,Wil06,Bro06}, this appears to be due to the truncation of
star formation in blue galaxies, and the role of mergers is poorly known.
\citet{vD05} and \citet{Bel06a,Bel06b}, using close galaxy pairs, conclude
that $L_\star$ red galaxies grow rapidly via mergers since $z=1$, while
\citet{Mas06}, using similar techniques, find that the merger rate of $4L_\star$
red galaxies is only $\sim 1\%~{\rm Gyr}^{-1}$.
Using the galaxy space density, \citet{Bro06} find that the stellar masses of
$4L_\star$ red galaxies grow by $\simeq 25\%$ since $z\sim 0.7$, while others
find no significant growth over similar redshift ranges
\citep[e.g.,][]{Bun06,Cim06,Cap06,Wake}.

There is an additional route to constraining the evolution of galaxies, which
is to use their clustering properties.  Building upon the theoretically
understood evolution of the dark matter halo population, we can obtain
complementary constraints which bypass the model dependence of stellar
evolution or merger times as a function of projected distance.
We illustrate this approach in this {\it Letter\/}, presenting
evidence from the evolution of their clustering that luminous red galaxies
undergo merging or disruption between $z\sim 0.9$ and $z\sim 0.5$. 

\section{The observational sample}

We use galaxies in the $9\,{\rm deg^2}$ Bo\"{o}tes field, which has been
imaged in the optical and infrared by the NOAO Deep Wide-Field
\citep[NDWFS;][]{JanDey} and {\it Spitzer\/} IRAC Shallow Surveys
\citep{Eis04}.  We use a luminous ($>1.6\,L_\star$) subset of the Bo\"{o}tes
red galaxy sample which was selected from the imaging using empirical
photometric redshifts and an evolving restframe $U-V$ color criterion
\citep{Bro06}.
This subset includes galaxy samples in three redshift slices: $0.4<z<0.6$, 
$0.6<z<0.8$ and $0.8<z<1.0$ with comoving volumes of $2.4$, $3.5$ and 
$5.2\times 10^6\,(h^{-1}{\rm Mpc})^3$ respectively, and at each redshift,
galaxies are selected to be above a luminosity threshold such that
the sample has a constant comoving number density
($\bar{n}=10^{-3}\,h^{3}{\rm Mpc}^{-3}$).
Our results are based on the observed evolution of the angular clustering of
these samples, containing a few thousand galaxies each.
We transform from models of the spatial clustering to angular clustering using
a redshift distribution model which accounts for the small measured
uncertainties of the photometric redshifts ($\sigma_z\lesssim 0.05$).
We describe the clustering measurements and modeling in detail
in \citet{Clustering}.

\section{Modeling galaxy clustering}

\begin{figure}
\begin{center}
\resizebox{3.3in}{!}{\includegraphics{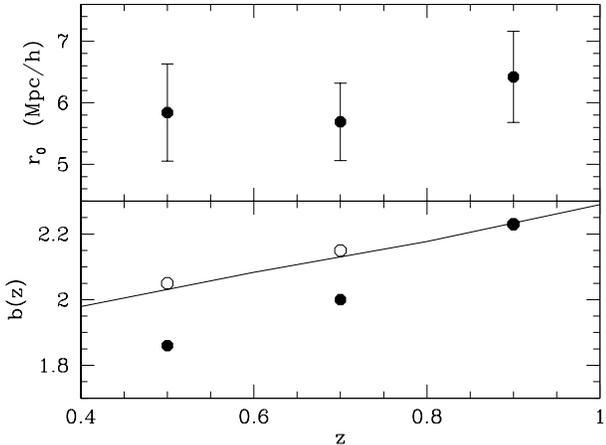}}
\end{center}
\caption{(Top) The correlation length, $r_0$, from a power-law fit for the
sample as a function of $z$ from \protect\citet{Clustering}.
(Bottom) The evolution of the large-scale bias, assuming a dark matter
power spectrum with $\sigma_8=0.8$.  The solid line is
$b(z)=1+D^{-1}\left[ b(0)-1\right]$ (see text),
the open circles are measured from our ``passive mocks'' and the solid circles
are from the mocks which best fit $w(\theta)$ at each redshift.}
\label{fig:b_vs_z}
\end{figure}

Our galaxy samples have been selected to have constant $\bar{n}$.
We find that their clustering evolves very little, with
$\xi(6\,h^{-1}{\rm Mpc},z)\simeq 1$.
We begin with an analytic argument to show that these galaxies cannot be
undergoing a common luminosity evolution history, such as pure passive
evolution, without any mergers\footnote{M.W., M.B.~and A.D.~thank Ravi Sheth
for discussions at the Aspen Center for Physics which generated this
argument.}.
In such a scenario each galaxy preserves its identity and no galaxy leaves
or enters the sample.
If we assume that galaxies and mass follow the same velocity field, and retain
their identities, then the continuity equation in the linear regime requires
that $\dot{\delta}_{\rm gal}=\dot{\delta}_{\rm m}$ \citep{Pee80}.  If we define
$\delta_{\rm gal}(z)=b(z)\delta_{\rm m}(z)$ and the growth function
$D(z)\equiv\delta_{\rm m}(z)/\delta_{\rm m}(0)$ then
$b(z)=1+D^{-1}(z)\left[ b(0)-1 \right]$ \citep{Fry96}.  
As shown in Figure~\ref{fig:b_vs_z},
this prediction is in good agreement with our numerical simulations with 
passive evolution (see \S~\ref{sec:passive}).
Assuming scale-independent, deterministic biasing therefore predicts evolution
in $\xi$ which is not in agreement with the observations.
In fact, we find that passive evolution cannot fit the trend of the central
values of the clustering strength for any plausible cosmology.

To go further we need a way of connecting the galaxies we observe with
the host dark matter halos whose evolution theory predicts.
The halo model (see e.g.~\citealt{CooShe} for a review) has provided us with
such a physically informative and flexible means of describing galaxy bias.
The key insight is that an accurate prediction of galaxy clustering requires
a knowledge of the occupation distribution of objects in halos (the HOD) and
their spatial distribution.  In combination with ingredients from N-body
simulations a specified HOD makes strong predictions about a wide array
of galaxy clustering statistics.
The formalism thus allows us to use observations of galaxy clustering to
constrain the connection between galaxies and their host dark matter halos
at each $z$, and in particular to show that the luminous galaxies in
the NDWFS have undergone significant merging or disruption between
$z\simeq 0.9$ and $z\simeq 0.5$ by comparing the HOD inferred from the
$z\simeq 0.5$ clustering data to one which is passively evolved from the
inferred HOD at $z\simeq 0.9$.

\subsection{Simulations and mock catalogs}

Our modeling of galaxy clustering is based on mock catalogs constructed
within the HOD framework by populating halos in a cosmological N-body
simulation.  We use a high resolution simulation of a $\Lambda$CDM cosmology
($\Omega_M=0.25=1-\Omega_\Lambda$, $\Omega_B=0.043$, $h=0.72$, $n_s=0.97$
and $\sigma_8=0.8$).
The linear theory power spectrum was computed by evolution of the coupled
Einstein, fluid and Boltzmann equations using the code described in
\citet{Boltz}.  This code agrees well with {\sl CMBfast\/} \citep{CMBfast},
see e.g.~\citet{SSWZ}.
The simulation employed $1024^3$ particles of mass
$8\times 10^{9}\,h^{-1}M_\odot$ in a periodic cube of side $500\,h^{-1}$Mpc
using a {\sl TreePM\/} code \citep{TreePM}.
The Plummer equivalent softening was $18\,h^{-1}$kpc (comoving).

For each output we generate a catalog of halos using the Friends-of-Friends
(FoF) algorithm \citep{DEFW} with a linking length of $0.168\times$
the mean inter-particle spacing.
This procedure partitions the particles into equivalence classes, by linking
together all particle pairs separated by less than a distance $b$.
The halos correspond roughly to particles with
$\rho>3/(2\pi b^3)\simeq 100$ times the background density.
Our mass definition uses the sum of the particle masses in the halo,
however to obtain better correspondence between our definition of halo mass
and that implicitly defined by the mass functions of \citet{SheTor} and
\citet{JFWCCECY} we rescaled the masses by
$M/M_{\rm fof}=1+0.01\left( \ln M_{\rm fof}-23.5 \right)$ where $M_{\rm fof}$
is the FoF mass in units of $h^{-1}M_\odot$.  With this redefinition the mass
function in the simulation lies between those of \citet{SheTor} and
\citet{JFWCCECY}, differing from them by less than 10\% in the mass range
of interest.

To make mock catalogs we use a halo model which distinguishes between central
and satellite galaxies. We choose a mean occupancy of halos:
$N(M)\equiv\left\langle N_{\rm gal}(M_{\rm halo})\right\rangle$.
Each halo either hosts a central galaxy or does not, while the number of
satellites is Poisson distributed about a mean $N_{\rm sat}$.
With the luminosity-threshold samples, we parameterize 
$N(M)=N_{\rm cen}+N_{\rm sat}$ with 4 parameters \citep[e.g.][]{Zheng05}
\begin{equation}
  N_{\rm cen}(M) = \frac{1}{2}
  \ {\rm erfc}\left[\frac{\ln(M_{\rm cut}/M)}{\sqrt{2}\sigma}\right]
\label{eqn:ncen}
\end{equation}
and
\begin{equation}
  N_{\rm sat}(M) = \left(\frac{M-M_{\rm cut}}{M_1}\right)^\alpha
\label{eqn:nsat}
\end{equation}
for $M>M_{\rm cut}$ and zero otherwise.  Different functional forms have
been proposed in the literature, but the current form is flexible enough
for our purposes.  Including a different low mass roll-off in the satellite
term, following \citet{TWZZ} and \citet{CWK}, does not alter our basic
conclusions.

Given an HOD and the halo catalogs we can produce a mock catalog in one of
two ways.  Central galaxies always live at the minimum of the halo potential.
We either place satellite galaxies assuming an NFW profile \citep{NFW} with
a concentration-mass relation fit to the halos in the simulation or
anoint $n_{\rm sat}$ dark matter particles, chosen at random, as
galaxies.  The two methods produce very similar, though not identical,
clustering, with the biggest differences on Mpc scales.  An analytic model
\citep[described in][]{Zheng04,TWZZ} also produces very similar results.
The differences between the methods are smaller than the observational errors,
so we shall neglect them henceforth.

\begin{figure}
\begin{center}
\resizebox{3.3in}{!}{\includegraphics{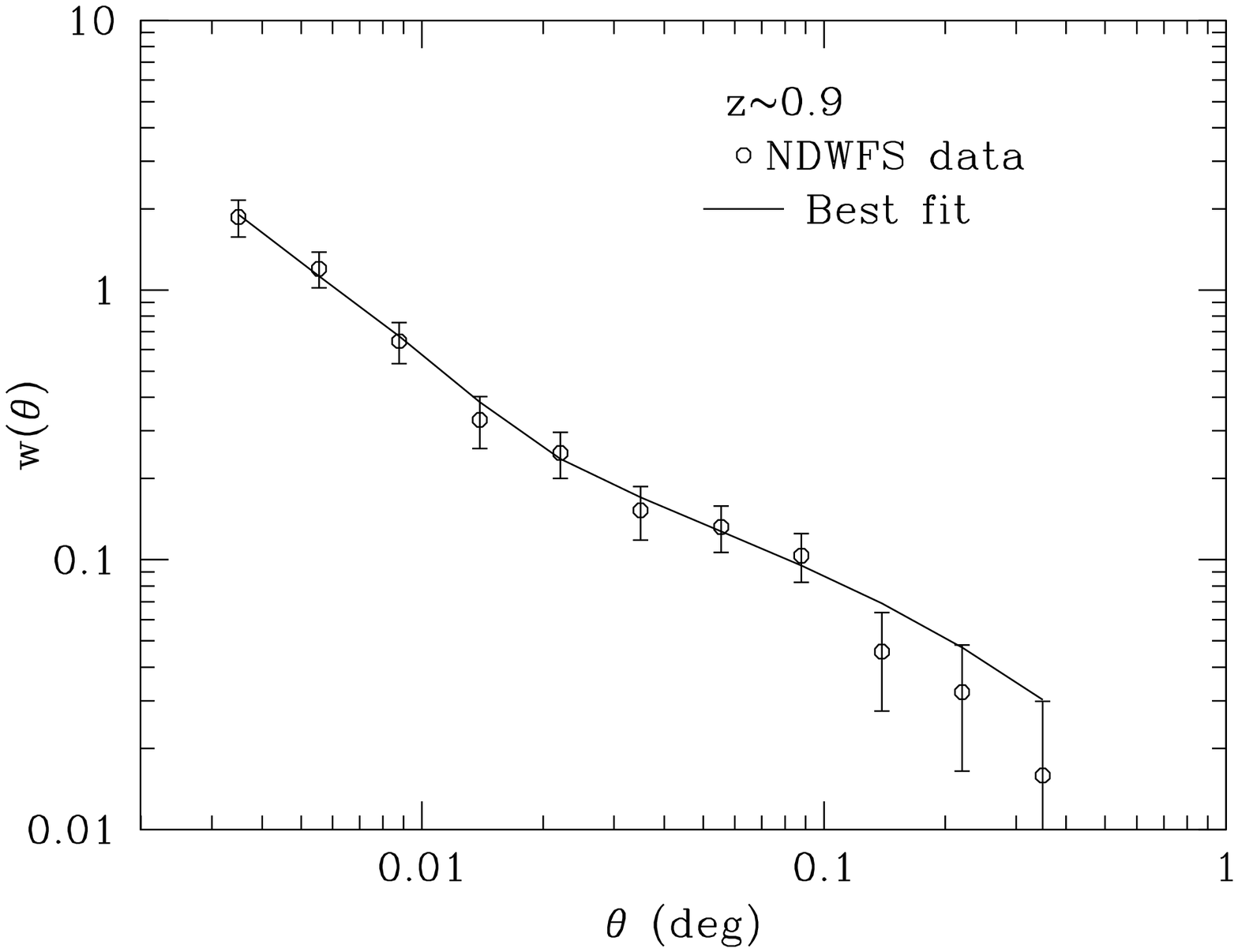}}
\resizebox{3.3in}{!}{\includegraphics{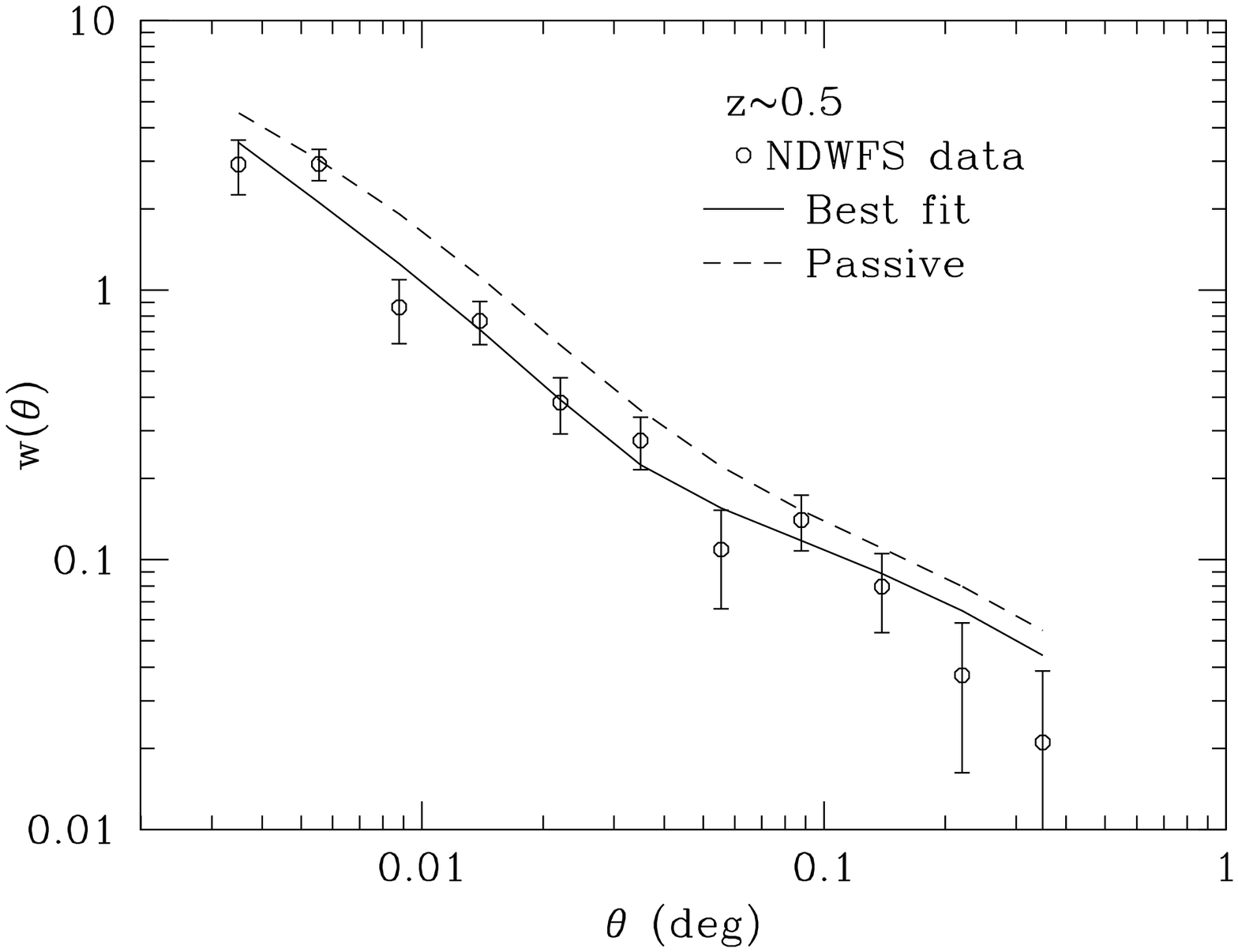}}
\end{center}
\caption{The angular correlation function, $w(\theta)$, for the $0.4<z<0.6$
and $0.8<z<1.0$ slices.  Open circles with error bars represent the Bo\"{o}tes
data, the solid line is the best fitting HOD model prediction and (in the
lower panel) the dashed line is the prediction of the best fitting
$z\simeq 0.9$ model passively evolved to $z\simeq 0.5$.}
\label{fig:wtheta}
\end{figure}

\subsection{Comparing with data} \label{sec:fits}

{}From the model galaxy positions we compute $\xi(r)$ in real space by direct
pair counting in the periodic box for separations $<20\,h^{-1}$Mpc.
Beyond $20\,h^{-1}$Mpc we extrapolate assuming a constant bias.
The redshift distribution is used to convert $\xi(r)$ into $w(\theta)$ using
Eq.~(50) of \citet{Sim06}, giving the predicted clustering for any set of HOD
parameters.  We fit to the data assuming Gaussian errors with the covariance
matrices of \citet{Clustering}, and assume a 5\% error on the number density
of galaxies.
Figure~\ref{fig:wtheta} compares the best fitting HOD model predictions
to the data at $z\simeq 0.9$ and 0.5.

In order to propagate the observational errors into uncertainties in the
HOD (Eqs.~\ref{eqn:ncen},\ref{eqn:nsat}) we used a Markov chain Monte-Carlo
method \citep[e.g.][]{MCMC} as detailed in \citet{Clustering}.
We found that the data were unable to rule out models with $\sigma\gg 1$
and $\alpha\ll 1$, which we regard as unlikely for large red galaxies, so
we impose a prior which penalizes $\sigma>1$ and $\alpha\simeq 0$.
Tests indicate that surveys twice as large would not need this prior,
though with this prior even our chains converged well.
Because the mock catalog generation using NFW profiles is very fast and
requires little memory we use this to generate the chains.

\begin{figure}
\begin{center}
\resizebox{3.3in}{!}{\includegraphics{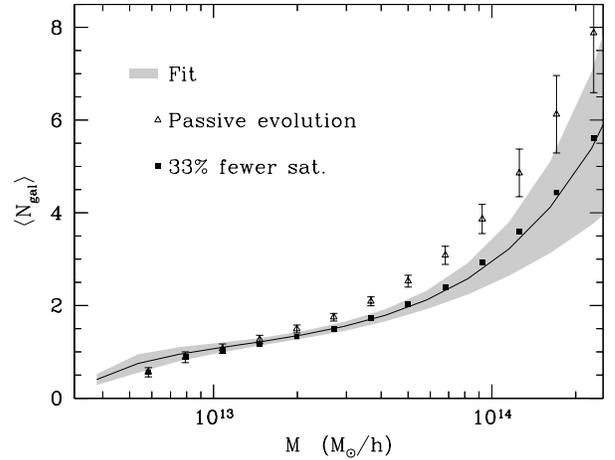}}
\end{center}
\caption{HODs for the $z\simeq 0.5$ sample.  The shaded area indicates the
mean and standard deviation in the HOD from the Markov chains, fit to the
$w(\theta)$ data at $z\simeq 0.5$.  The mass scale has been increased by
7\% as described in the text.  Open triangles indicate the HOD from the
models that fit the $z\simeq 0.9$ data evolved to $z\simeq 0.5$ by tracking 
particles. Squares assume 33\% of the satellites in the passively evolved 
mocks disappear between $z\simeq 0.9$ and $z\simeq 0.5$.}
\label{fig:hod}
\end{figure}

\subsection{Passive evolution} \label{sec:passive}

For a subset of 250 of the models at $z\simeq 0.9$ we use the particle-based
method to produce mock catalogs which we passively evolve to $z\simeq 0.5$
simply by tracking the particles based on their ID.  The positions and halo
memberships of these tagged particles are followed.
We expect the HOD to be more robust than the small-scale clustering for
these particles\footnote{The small-scale clustering depends on the evolution
of the subhalos inside of the host halo, and due to finite force and mass
resolution these may not be correctly modeled in massive halos.}, but we show
the latter in Figure~\ref{fig:wtheta} for completeness.
Looking at the difference in the HODs is also more informative, and gives us
a clue as to what physics may be missing from pure passive evolution.

\section{Discussion}

A comparison of the HOD of the passively evolved samples with the HODs which
best fit the $z\simeq0.5$ data indicates that evolution produces too many 
galaxies in high mass halos, as shown in Figure~\ref{fig:hod}.
A similar conclusion can be reached by comparing the clustering of the
passively evolved models to the data in Figure~\ref{fig:wtheta} --- the
excess clustering on small scales from passive models clearly indicates that
the models overpredict galaxy pairs within the same halo, i.e., they predict
too many satellite galaxies in high mass halos.
Another indication of the excess is that the satellite fraction in the
passively evolved models is $0.24\pm 0.02$ while that in the best fitting
models is $0.18\pm 0.02$.

There must be some physical process which reduces the number of galaxies
in massive halos, and the most natural candidates are dynamical friction
and tidal stripping which act to merge massive satellites with the central
galaxy or disrupt them.
Based on Figure~\ref{fig:hod}, at high halo masses $\sim 1/3$ of the 
satellites in the passively evolved catalogs must have disappeared by
$z\simeq 0.5$.  We caution that our calculation ignores sources and sinks.
If galaxies can enter or leave the sample due to rapid evolution of their
star formation rates, interpretting the evolution of galaxy clustering is
non-trivial.  However, $z<1$ blue galaxies with masses comparable to the
most massive red galaxies are rare, and (at least at $z\simeq 0$) red
galaxies have little cold gas to fuel renewed bursts of star formation.

There is a subtlety to bear in mind with these statistics.
Merging results in a small fraction of $z\simeq 0.9$ galaxies disappearing
by $z\simeq 0.5$.  So we should really compare the passively evolved catalog
(with $\bar{n}=10^{-3}\,h^3{\rm Mpc}^{-3}$) to a fitted one with a slightly
lower number density -- the ``true'' $z\simeq 0.5$ descendants of the
$z\simeq 0.9$ galaxies.
The cut-off mass scale for this catalog would be slightly larger than our
fits and we should shift the mass definition in the HOD accordingly.
We estimate the size of this shift, by considering the evolution of the
central galaxies, to be $\approx 7\%$ and in Figure~\ref{fig:hod} we increase
the mass scale of the HOD inferred from the $z\simeq 0.5$ data accordingly
(though this shift does not change our conclusions).

To estimate the impact of the disappeared satellites on the growth of the 
central galaxy or the boost of the intracluster light (ICL) in the halo,
we make the following simple model. 
By matching the number density of halos above mass $M$ with that of galaxies
above luminosity $L$ (we use the $B$-band LFs of \citealt{Bro06}),
we can relate the central galaxy luminosity to halo mass (the luminous end
of the LF is dominated by central galaxies as shown in \citealt{Zheng05}).
For our sample, at $z\simeq 0.5$ a halo of $5\times 10^{13}\,h^{-1}M_\odot$
on average hosts a central galaxy of $\approx 5\,L_\star$, with
$L_{\rm cen}\propto M^{0.36}$ for massive halos.
If we also assume that the satellite LF has the same shape as the global LF,
we can use the fitted HODs to find the number of satellites and integrate
to measure their total light.
We find that in such halos satellites on average have a total lumnosity of
$2.6\,L_\star$, so if this is what is left after $1/3$ of the satellites
disappeared, the satellites would have contributed $\approx 25\%$ of the
current stellar mass to the central galaxy or a similar mass to the ICL.
For halos of $10^{14}\,h^{-1}M_\odot$ the contribution is $\approx 40\%$.
These numbers are {\it upper bounds\/}, since satellites may have lost
just enough mass to leave our sample.
Note that the stellar material resides in the inner regions of the halo,
where the potential well is the deepest, and is the last material to be
disrupted.  An ``average'' satellite would need to lose 40\% of its stellar
mass, or decrease its surface brightness, to drop out of the sample.
Analyzing samples with lower luminosities would help us constrain this.

We are unable, without further modeling, to differentiate between satellites
fueling the growth of the central galaxy or the ICL, which at lower
redshift comprises $\approx 5-10\%$ of the stellar mass in groups and
clusters \citep{Gon,Zib,Agu}.
We can, however, form an $L_{\rm cen}-M$ relation at $z\simeq 0.9$ as above.
By matching progenitor halos to their descendants at $z\simeq 0.5$ and
assuming $0.48$ $B$-band magnitudes of fading we find $\approx 10\%$ growth
in the stellar mass of the central galaxy between $z\simeq 0.9$ and $0.5$.
This would suggest the satellites also build an ICL component (the total
stellar mass in the disappeared satellites, if it all ended up as ICL,  
constitute $\sim 15\%$ of the total halo luminosity above the threshold).
However this result relies on our assumption of a uniform LF and on stellar
population evolution predictions, and needs to be constrained by observations
of the ICL.
If we argue that there is little extended light at high $z$, either the
satellites may have lost just enough mass to leave the sample or the
``extra'' mass would accrete onto the central galaxy.
We regard this dichotomy as an open question requiring further investigation.

Our conclusions are necessarily tentative due to the limited volume of the
NDWFS Bo\"{o}tes survey, which does not probe the mass function above
$10^{14}\,h^{-1}M_\odot$ well.
There are several areas where more or different data would be beneficial.
Tests using models of larger surveys\footnote{The wider area survey need
not be as deep as NDWFS.} indicate that doubling the survey volume removes
the islands of parameter space which we have excluded with priors and
shrinks the errors on the HOD parameters by $\approx\sqrt{2}$.
A measurement of the space density of groups richer than several members
would shrink the errors on the high mass end of the HOD dramatically,
but would require more volume than we have at present to contain a
representative sample of rich groups.
We also investigated the dependence of our results upon cosmology using
similar simulations with different parameters.  Our results remain robust
within the currently allowed range of models.

This preliminary investigation shows the power of clustering measures
to inform questions of the formation and evolution of galaxies.  We find
evidence for evolution in the red galaxy HOD very different than that
predicted by pure passive evolution models.
Our result is largely independent of models of red galaxy stellar populations
and estimates of dynamical friction scales.
With more data from the NDWFS and future surveys we hope to be able to trace
in detail the formation history of the most massive galaxies in the Universe.

\begin{acknowledgments}
We would like to thank Ravi Sheth for conversations and Jerry Ostriker for
comments on an early draft.  We thank both Charlie Conroy and David Weinberg
for emphasizing the importance of satellite disruption.  We thank the Aspen
Center for Physics, where this work was begun, for their hospitality.
M.W. thanks the Galileo Galilei Institute for Theoretical Physics for
their hospitality and the INFN for partial support during the completion
of this work.
This work is based in part on observations from the National Optical
Astronomy Observatory, operated by AURA, Inc.~under a cooperative agreement
with the NSF, and the Spitzer Space Telescope.
The simulations were performed on the supercomputers at the National
Energy Research Scientific Computing center.
MW was supported by NASA. Z.Z.~acknowledges the support of NASA through a
Hubble Fellowship grant
HF-01181.01-A awarded by the Space Telescope Science Institute, which
is operated by the Association of Universities for Research in Astronomy,
Inc., for NASA, under contract NAS 5-26555.

\end{acknowledgments}


\begin{thebibliography}{99}

\bibitem[{{Aguerri et al.}(2006)}]{Agu}
Aguerri J.A.L., Castro-Rodriguez N., Napolitano N., Arnaboldi M.,
  Gerhard O., 2006, A\&A, 457, 771

\bibitem[{{Bell et al.}(2004)}]{Bel04}
Bell E.F., et al., 2004, \apj, 608, 752

\bibitem[Bell et al.(2006a)]{Bel06a} 
Bell E.F., et al., 2006a, \apj, 640, 241 

\bibitem[Bell et al.(2006b)]{Bel06b} 
Bell E.F., et al., 2006b, \apj, 652, 270

\bibitem[{{Bower, Lucey \& Ellis}(1992)}]{BowLucEll}
Bower R.G., Lucey J.R., Ellis R.S., 1992, \mnras, 254, 601

%\bibitem[{{Brown et al.}(2003)}]{Bro03}
%Brown M.J.I., et al., 2003, \apj, 597, 225

\bibitem[{{Brown et al.}(2006a)}]{Bro06}
Brown M.J.I., et al., 2006a, \apj, in press [astro-ph/0609584]

\bibitem[{{Brown et al.}(2006b)}]{Clustering}
Brown M.J.I., et al., 2006b, in preparation.

\bibitem[{{Bundy et al.}(2006)}]{Bun06}
Bundy K., et al., 2006, \apj, in press [astro-ph/0512465]

\bibitem[{{Cimatti, Daddi \& Renzini}(2006)}]{Cim06}
Cimatti A., Daddi E., Renzini A., 2006, A\&A, 453, 29

\bibitem[{{Caputi et al.}(2006)}]{Cap06}
Caputi K.I., et al., 2006, \mnras, 366, 609

\bibitem[{{Conroy, Wechsler \& Kravtsov}(2006)}]{CWK}
Conroy C., Wechsler R.H., Kravtsov A.V., 2006, \apj, in press [astro-ph/0512234]

\bibitem[Cool et al.(2006)]{Coo06}
Cool, R.~J., et al., 2006, \aj, 131, 736 

\bibitem[{{Cooray \& Sheth}(2002)}]{CooShe}
Cooray, A., Sheth, R., 2002, Phys. Rep., 372, 1 [astro-ph/0206508]

\bibitem[{{Davis et al.}(1985)}]{DEFW}
Davis M., Efstathiou G., Frenk C.S., White S.D.M., 1985, ApJ, 292, 371

\bibitem[Eisenhardt et al.(2004)]{Eis04} 
Eisenhardt, P.~R., et al., 2004, \apjs, 154, 4

%\bibitem[{{Faber et al.}(2006)}]{Fab06}
%Faber S., et al., 2006, \apj, in press [astro-ph/0506044]

\bibitem[Fry(1996)]{Fry96}
Fry, J.~N.\ 1996, \apjl, 461, L65

%\bibitem[{{Glazebrook et al.}(2004)}]{GDDS}
%Glazebrook K., et al., 2004, Nature, 430, 181

\bibitem[{{Gonzalez et al.}(2000)}]{Gon}
Gonzalez A.H., Zabludoff A.I., Zaritsky D., Dalcanton J., 2000,
  \apj, 536, 561

%\bibitem[{{Heymans et al.}(2006)}]{Hey06}
%Heymans C., et al., 2006, \mnras, 371, L60

\bibitem[{{Jenkins et al.}(2000)}]{JFWCCECY}
Jenkins A., Frenk C.S., White S.D.M., Colberg J.M., Cole S., Evrard A.E.,
Couchman H.M.P., Yoshida N., 2001, MNRAS, 321, 372

\bibitem[{{Gilks, Richardson \& Spiegelhalter}(1996)}]{MCMC}
Gilks W.R., Richardson S.R., Spiegelhalter D.J., ``Markov chain Monte
Carlo in practice'', Chapman \& Hall (Florida, 1996)

\bibitem[{{Jannuzi \& Dey}(1999)}]{JanDey}
Jannuzi B.T., Dey A., 1999, in ASP Conf. Ser. 191, Photometric
redshifts and high redshift galaxies, ed. R.J. Weymann, L.J. Storrie-Lombardi,
M. Sawicki \& R.J. Brunner (San Francisco, ASP), 111

%\bibitem[{{Jannuzi et al.}(2007)}]{Jan07}
%Jannuzi B.T., et al., 2007, in preparation

%\bibitem[{{Kravtsov et al.}(2004)}]{KBWKGAP}
%Kravtsov A.V., et al., 2004, \apj, 609, 35.

\bibitem[{{Masjedi et al.}(2006)}]{Mas06}
Masjedi M., et al., 2006, \apj, 644, 54

\bibitem[{{Navarro, Frenk \& White}(1996)}]{NFW}
Navarro, J., Frenk, C., White, S.D.M., 1996, \apj, 462, 563 

\bibitem[{{Peebles}(1980)}]{Pee80}
Peebles P.J.E., 1980, ``The large scale-structure of the universe'',
(Princeton, New Jersey).

\bibitem[{{Seljak \& Zaldarriaga}(1996)}]{CMBfast}
Seljak U., Zaldarriaga M., 1996, \apj, 469, 437.

\bibitem[{{Seljak et al.}(2003)}]{SSWZ}
Seljak U., Sugiyama N., White M., Zaldarriaga M., 2003,
  Phys. Rev. D68, 83507.

\bibitem[{{Sheth \& Tormen}(1999)}]{SheTor}
Sheth R., Tormen G., 1999, \mnras, 308, 119

\bibitem[{{Simon}(2006)}]{Sim06}
Simon P., 2006, A\&A, in press [astro-ph/0609165]

\bibitem[{{Trager et al.}(2000)}]{Tra00}
Trager S.C., Faber S.M., Worthey G., Gonzalez J.J., 2000, \aj, 120, 165

\bibitem[{{Tinker et al.}(2005)}]{TWZZ}
Tinker J.L., Weinberg D.H., Zheng Z., Zehavi I., 2005, \apj, 631, 41

\bibitem[{{van Dokkum}(2005)}]{vD05}
van Dokkum P.G., 2005, \aj, 130, 2647

\bibitem[{{Wake et al.}(2006)}]{Wake}
Wake D.A., et al., 2006, \mnras, 372, 537

\bibitem[{{White}(2002)}]{TreePM}
White M., 2002, ApJS, 579, 16

\bibitem[{{White \& Scott}(1995)}]{Boltz}
White M, Scott D, 1995, \apj, 459, 415

\bibitem[{{Willmer et al.}(2006)}]{Wil06}
Willmer C.N.A., et al., 2006, \apj, 647, 853

\bibitem[{{Zheng}(2004)}]{Zheng04}
Zheng Z., 2004, \apj, 610, 61

\bibitem[{{Zheng et al.}(2005)}]{Zheng05}
Zheng Z., et al., 2005, \apj, 633, 791

\bibitem[{{Zibetti et al.}(2005)}]{Zib}
Zibetti S., White S.D.M., Schneider D.P., Brinkmann J., 2005,
  \mnras, 358, 949

\bibliographystyle{apj}  
\end{thebibliography}
\end{document}